# The study of periphery uniqueness and balance in ecological networks

Shipeng Xu

College of Agricultural and Biological Sciences, Dali University, Dali 671003, China

xsp@stu.dali.edu.cn

https://orcid.org/0009-0002-0921-5318

## Abstract:

The study of ecological networks is pivotal in modern conservation biology, addressing pressing issues such as habitat fragmentation and biodiversity loss, particularly in geographically complex regions. These networks, comprising ecological corridors, sources, and nodes, are essential for facilitating species movement, genetic exchange, and the overall functioning of ecosystems. Traditional network analysis has primarily focused on central nodes, overlooking the critical role of peripheral nodes that form the interface with the external environment. Peripheral nodes, though often less connected, are vital for the network's response to environmental changes, acting as buffers or filters and influencing the network's resilience and adaptability. This research introduces the innovative Periphery Analysis Model (PAM), specifically designed to examine the periphery of ecological networks. PAM is the first methodological framework dedicated to periphery analysis, incorporating theories and methodologies from graph theory, complex network analysis, and landscape ecology. The model is centered around two principal indices: the Periphery Uniqueness Index (PuI) and the Periphery Balance Index (PbI), which quantify the distinctive attributes and equilibrium of peripheral nodes within the ecological network. PAM also includes derived indices for a more nuanced understanding of the periphery's role and influence. The application of PAM promises to illuminate the characteristics defining the periphery of ecological networks, revealing the intricate interdependencies and interactions between peripheral nodes and the network as a whole. This model is expected to enhance our understanding of the deeper structural features of ecological networks, offering insights into the maintenance of biodiversity, ecological connectivity, and ecosystem health. The study advocates for the integration of PAM in conservation strategies, highlighting its potential to inform policy and management decisions for preserving ecosystem stability and resilience in the face of environmental challenges.

# 1. Introduction

Ecological networks(ENs) construction is a crucial approach in modern conservation biology, addressing challenges like habitat fragmentation and biodiversity loss, particularly in regions with complex geographical environments (Xiao et al., 2022). These networks, consisting of ecological corridors, sources, and nodes, facilitate species movement, genetic exchange, and overall ecosystem functioning (Geng et al., 2023). Methods like Morphological Spatial Pattern Analysis (MSPA) for identifying ecological sources and the Minimum Cumulative Resistance (MCR) model for determining ecological corridors are essential tools in this process (Z. Yang et al., 2023; X. Zhou et al., 2023).MSPA plays a key role in ENs planning by identifying and mapping crucial habitat areas known as "ecological sources," which are vital for maintaining biodiversity. These sources serve as the building blocks of ecological networks, providing core areas for species to thrive (Chen et al., 2023). The MCR model is another essential method, identifying the most effective routes for ecological corridors that connect different ecological sources. These corridors are crucial for species movement, allowing them to migrate, disperse, and interact across fragmented landscapes (D. Zhou & Song, 2021).Additionally, the gravity model is also employed in ENs planning. This model assesses the interaction strength between different ecological sources, aiding in the prioritization and optimization of corridors based on the potential for species movement and interaction (C. Yang et al., 2022).

Research on Ecological Networks (ENs) has primarily concentrated on developing methods for the construction and subsequent evaluation of these networks, with a particular emphasis on pinpointing the central nodes that play a crucial role within the network's framework. The identification of these key nodes is achieved through the computation of various indices, including node centrality, which measures the influence of a node within the network; betweenness centrality, which gauges a node's importance in connecting other nodes; and clustering coefficients, which indicate the degree of interconnection within a node's neighborhood (Z. Wang, et al., 2023). These central nodes are deemed to possess a higher level of importance within the network, exerting a substantial influence on the network's overall stability and robustness, as they often serve as critical hubs for the flow of information, resources, and interactions (Li, et al., 2021). Despite the significant progress made in this area, there exists a notable gap in the literature when it comes to the study of peripheral nodes within ecological networks. Peripheral nodes, while often overlooked, are an essential component of the network, as they form the outer layers that interface with the external environment and facilitate the exchange of materials, energy, and organisms between the network and its surroundings. The properties and dynamics of these peripheral nodes are instrumental in shaping the network's response to environmental changes and stressors, as they can act as buffers or filters, influencing the resilience and adaptability of the network to various disturbances. Therefore, a comprehensive understanding of the periphery's characteristics and dynamics is crucial for gaining insights into the deeper structural features of the entire network, as they offer an alternative perspective that complements the focus on central nodes. By examining the peripheral nodes, researchers can uncover patterns and processes that may not be evident when concentrating solely on the network's core, such as the role of peripheral nodes in maintaining biodiversity, facilitating ecological connectivity, and promoting the overall health and functioning of the ecosystem.

In addressing the aforementioned research gaps, our study specifically focuses on the Ecological Networks Periphery (ENP) as the central object of investigation and proposes an innovative "Periphery Analysis" model tailored for the examination of ecological networks. This "Periphery Analysis" model represents a groundbreaking approach by incorporating a rich tapestry of theories and methodologies from graph theory, complex network analysis, and landscape ecology. It stands as the inaugural methodological framework explicitly dedicated to the periphery analysis within the intricate fabric of complex ecological networks. The "Periphery Analysis" model is meticulously structured around two principal indices that serve as the cornerstone of its analytical framework: the Periphery Uniqueness Index and the Periphery Balance Index. These indices are meticulously crafted to quantify and evaluate the distinctive attributes of peripheral nodes and their equilibrium within the broader context of the ecological network. Furthermore, the model expands its analytical prowess through the inclusion of a suite of derived indices. These derived indices are calculated based on a nuanced interplay of various network characteristics, thereby offering a more comprehensive understanding of the periphery's role and influence within the ecological network. The utilization of the "Periphery Analysis" model is poised to provide a profound insight into the characteristics that define the periphery of ecological networks. It will illuminate the nuanced interdependencies and interactions that exist between peripheral nodes and the rest of the network. Moreover, by delving into the periphery's attributes, this model will shed light on the structural intricacies and underlying dynamics that contribute to the overall stability and resilience of the ecological network.

## 2. Methods

### 2.1 Definition and identification of the periphery

The definition of the network periphery is associated with the core-periphery structure (Elliott et al., 2020). In this structure, the network is divided into two main parts: the core and the periphery. The periphery typically refers to a collection of nodes that are connected to the core nodes but have relatively few connections with each other. In the core-periphery structure, peripheral nodes have the following characteristics: (1) Connections between peripheral nodes are relatively sparse. (2) Peripheral nodes are primarily connected to the rest of the network through the core nodes. (3) The periphery can be seen as the outer part of the network, where the internal connections of nodes are not as tight as those in the core area. Based on the core-periphery structure, there are two algorithms that can identify the network periphery.

**2.1.1 Algorithm based on random walks**

This method identifies the core-periphery structure by analyzing the behavior of random walkers. The core-periphery structure is revealed by considering the actions of random walkers within the network. In this approach, peripheral nodes are identified by calculating the persistence probability of nodes, which is the probability that a random walker remains on a particular node for a given amount of time (Rossa et al., 2013). The algorithm constructs the core-periphery structure through an iterative process, starting with the most weakly connected nodes and progressively adding those most likely to belong to the core. The mathematical formula of the algorithm is as follows:

$$a_i = 1 - \left(1 - p_{ij}\right)^{N_i}$$

Here, $a_i$ is the persistence probability of node $i$, $p_{ij}$ is the probability of a connection existing between nodes $i$ and $j$, and $N_i$ is the number of neighbors of node $i$.

**2.1.2 Algorithm based on core decomposition**

This method identifies the core-periphery structure by iteratively removing the connections of nodes. In k-shell decomposition, nodes in the network are assigned to different "shells" based on the "strength" of their connections. Each shell layer represents a connectivity level of the nodes in the network, with nodes in the outer layers having fewer connections and those in the inner layers having more connections (Zhang et al., 2015). This method can reveal the hierarchical structure of nodes within the network, where core nodes are typically located in the inner layers and peripheral nodes are in the outer layers. Below is the pseudocode form of the k-shell decomposition algorithm:

```
function kShellDecomposition(G):
   initialize shellCounter to 1
   initialize shells to an empty list

   while there exists a node v in G with k-shell equal to shellCounter:
      remove v and all its incident edges from G
      for each neighbor u of v:
         if u is still present in G:
            decrement the k-shell of u by 1
   end while

   return shells
```

## 2.2 Periphery analysis model

Periphery Analysis Model (PAM) is a new ecological network analysis model proposed by us, focusing on two key metrics: Periphery Uniqueness (Pu) and Periphery Balance (Pb). Below are the definitions and calculation methods for these two fundamental measures.

### 2.2.1 Periphery uniqueness

Periphery uniqueness refers to the extent to which the characteristics of peripheral nodes in an ecological network differ from those of core nodes. The Periphery Uniqueness Index (PuI) is the primary quantitative measure of periphery uniqueness. The Periphery Uniqueness Index (PuI) is further divided into the Network Periphery Uniqueness Index (N-PuI) and the Ecologically Weighted Periphery Uniqueness Index (E-PuI). The N-PuI is the Periphery Uniqueness Index calculated solely based on network characteristics, while the E-PuI is the Periphery Uniqueness Index obtained by incorporating weights based on the ecological characteristics of the peripheral nodes, such as biodiversity indices.

The N-PuI can be expressed as:

$$N\text{-}PuI_i = \frac{1}{|N_{periphery}|} \sum_{i \in N_{periphery}} \left| d_i - \bar{d}_{core} \right|$$

Where, $d_i$ is the degree of node $i$, $\bar{d}_{core}$ is the average degree of core nodes, and $|N_{periphery}|$ is the number of peripheral nodes.

The E-PuI can be expressed as:

$$E\text{-}PuI_i = \sum_{e \in E} w_e \cdot f_e(i) \times \left| d_i - \bar{d}_{core} \right|$$

Where, $f_e(i)$ is the quantified value of ecological indicator $e$ for node $i$, $w_e$ is the weight of indicator $e$, $d_i$ is the degree of node $i$, and $\bar{d}_{core}$ is the average degree of core nodes.

**2.2.2 Periphery balance**

Periphery balance refers to the extent to which the characteristics of peripheral nodes in an ecological network are close to or connected with those of core nodes. The Periphery Balance Index (PbI) is the primary quantitative measure of periphery balance. The Periphery Balance Index (PbI) is divided into the Network Periphery Balance Index (N-PbI) and the Ecologically Weighted Periphery Balance Index (E-PbI). The N-PbI is the Periphery Balance Index calculated solely based on network characteristics, while the E-PbI is the Periphery Balance Index obtained by incorporating weights based on the ecological characteristics of the peripheral nodes, such as the Biodiversity Index. Periphery balance and periphery uniqueness are relative concepts, but their calculation methods differ.

The N-PbI can be expressed as:

$$N\text{-}PbI_i = \frac{1}{|N_{periphery}|} \sum_{i \in N_{periphery}} \left( \frac{1}{|N|} \sum_{j \in N} l_{ij} \right)^{-1} \times \frac{L}{\max_{i \in N_{periphery}} \left( \frac{1}{|N|} \sum_{j \in N} l_{ij} \right)}$$

Where, $l_{ij}$ represents the shortest path length from node $i$ to node $j$, and $L$ is the maximum possible shortest path length in the network.

The E-PbI can be expressed as:

$$E\text{-}PbI_i = \frac{1}{|N_{periphery}|} \sum_{i \in N_{periphery}} \left( w_S \cdot S_i + w_{Sim} \cdot \frac{1}{|N_{core}|} \sum_{j \in N_{core}} Sim_{eco}(i,j) \right)$$

Where, $S_i$ represents the number of shortest paths between node *i* and core nodes, $Sim_{eco}(i, j)$ represents the similarity of ecological characteristics between node *i* and core node *j*, and $w_S$ and $w_{Sim}$ are the weights for the number of shortest paths and ecological characteristic similarity, respectively.

**2.3 Comprehensive analysis**

After identifying the periphery of ecological networks using specific algorithms, further analysis of the ecological networks is conducted using the periphery analysis model. The four indices N-PuI, E-PuI, N-PbI, and E-PbI, collectively provide a comprehensive framework for analyzing the structural and ecological characteristics of nodes within ecological networks. These indices have significant practical implications for understanding the complex dynamics of ecological networks. They can help identify critical nodes that may be more vulnerable to environmental changes or disruptions, prioritize areas for conservation efforts, and inform strategies for managing and preserving ecosystem health. When combined, these indices offer a multifaceted approach to ecological network analysis. For instance, a node with high N-PuI and low N-PbI might be a unique but isolated species that could be at risk if not properly conserved. Conversely, a node with high E-PuI and E-PbI might represent an ecologically significant species that plays a crucial role in maintaining the integrity of the ecosystem. By considering both network and ecological perspectives, researchers can develop more targeted and effective conservation strategies that take into account the complex interplay of biotic and abiotic factors.

# 3. Discussion

**3.1 The ecological significance of peripheral nodes**

Peripheral Nodes in ecological networks refer to those nodes that are connected to the core nodes of the network but do not occupy a central position within the network. These nodes typically have a lower degree of connectivity, yet they play an indispensable role in the overall structure and function of the network. The presence of peripheral nodes not only increases the redundancy and robustness of the network but also significantly impacts the transmission of information and the flow of energy within the network. Peripheral nodes play a critical role in maintaining the stability of ecological networks. As they are usually connected to the core nodes, peripheral nodes can act as a buffer for the transmission of information and the flow of energy. In the face of external disturbances or internal changes, peripheral nodes can slow down the propagation of shockwaves through their lower connectivity, thereby providing a certain protective mechanism for the network. Moreover, the existence of peripheral nodes also helps to disperse risks within the network, preventing the failure of a single node from causing the collapse of the entire network. The diversity of an ecological network is an important indicator of the health and stability of an ecosystem. Peripheral nodes facilitate the exchange of biodiversity and the flow of genes by connecting different ecological regions or species communities. This exchange of diversity not only aids species in adapting to environmental changes but also holds significant importance for the long-term evolution and adaptation of ecosystems. The existence of peripheral nodes makes ecological networks more complex and dynamic, thereby enhancing the ecosystem's adaptive capacity and resilience to environmental changes. Furthermore, peripheral nodes play a key role in the evolution of ecological networks. Due to their freedom and flexibility, peripheral nodes can experiment with new behaviors and strategies without being constrained or influenced by the core nodes. In this way, peripheral nodes can bring new information and opportunities to the network, thus promoting its evolution and innovation.

**3.2 Comparison of peripheral identification algorithms**

In network analysis, identifying the core-periphery structure is crucial for understanding complex systems. Two key algorithms for this task are the random walk-based and core decomposition-based methods. The random walk-based approach uses the behavior of a random walker to assign coreness values to nodes, revealing the network's global structure and the role of each node. This method is versatile for various network types, including directed and weighted networks, and can highlight nodes that bridge or are peripheral to core activities. In contrast, the core decomposition-based algorithm targets directed networks, partitioning them into core and periphery based on edge direction and block model formulations. This approach is adept at detecting complex core-periphery structures, especially in networks where the direction of interactions is pivotal. However, it may be less adaptable to weighted networks and can be computationally intensive. Overall, the random walk-based method excels in providing a continuous coreness measure, while the core decomposition-based method offers detailed insights into directed network structures. The selection of an algorithm should be based on the network's characteristics and the research objectives. Future research could integrate both methods to enhance the understanding of network structures.

## 3.3 Advantages and limitations of the periphery analysis model

The Periphery Analysis Model (PAM) offers several advantages in the study of ecological networks. Firstly, it provides a focused examination of peripheral nodes, which are often neglected in traditional network analysis. By quantifying periphery uniqueness and balance, PAM allows for a more nuanced understanding of the roles these nodes play in network stability and ecological connectivity. Secondly, the model's incorporation of both network characteristics and ecological weights provides a comprehensive perspective that integrates topological and ecological insights. This dual focus can lead to more informed conservation strategies that consider the complex interplay between species and their environment. However, PAM also has its limitations. The model's reliance on specific algorithms for periphery identification means that its effectiveness is contingent on the accuracy of these algorithms. Additionally, the calculation of derived indices can be complex and resource-intensive, particularly for large and complex networks. Furthermore, PAM may struggle to capture the full dynamics of peripheral nodes, especially in networks with high temporal variability or those subject to frequent disturbances.

## 3.4 Outlook for priphery research in networks

Future research on network peripheries should aim to refine and expand upon the PAM. One direction is to develop more efficient algorithms for periphery identification that can handle dynamic networks and provide real-time insights. Another is to explore the integration of additional ecological data, such as species interactions and environmental conditions, to provide a more holistic view of peripheral nodes' roles. Additionally, research could focus on the application of PAM in various ecological contexts, from urban ecosystems to wilderness areas, to assess its versatility and applicability across different environmental settings. Moreover, the integration of PAM with other disciplines such as urban planning and public health could lead to innovative solutions for managing urban green spaces, mitigating the impacts of climate change, and enhancing the well-being of urban populations. For instance, identifying peripheral nodes in urban ecosystems could inform the creation of green corridors that not only support biodiversity but also improve air quality and provide recreational spaces for residents. Furthermore, the expansion of PAM to global and regional scales could contribute to the study of large-scale ecological patterns and the assessment of environmental changes on a broader level. By incorporating satellite imagery and geospatial data, researchers can monitor the dynamics of peripheral nodes over time and across different ecosystems, providing valuable information for international conservation efforts and environmental agreements.

# 4. Conclusions

In conclusion, the PAM emerges as a pivotal tool in the study of ecological networks, offering a novel perspective on the role and significance of peripheral nodes. By focusing on the uniqueness and balance of these nodes, PAM provides a comprehensive framework that integrates network topology with ecological attributes, leading to a deeper understanding of the structural and functional aspects of ecological networks. The model's application extends beyond purely ecological contexts, showing promise in interdisciplinary fields such as ecological-social systems, urban planning, and even information technology. As we move forward, the continued development and refinement of PAM will be essential in addressing the challenges posed by environmental changes and in promoting the sustainable management of ecosystems. The model's potential to inform policy and conservation strategies, as well as its ability to contribute to the broader goal of ecological sustainability, makes it a valuable asset in the arsenal of tools available to researchers and practitioners alike.